\documentclass{aa}
\usepackage{graphicx}

\begin{document}
\title{ XMM-Newton Detection of Nova Muscae 1991 in Quiescence. }

\author{ F. K. Sutaria \inst{1}
     \and  U. Kolb \inst{1}  
     \and P. Charles \inst{2} 
     \and J. P.  Osborne \inst{3}
     \and E. Kuulkers \inst{4} 
     \and J. Casares \inst{5}
     \and E. T. Harlaftis \inst{6} 
     \and T. Shahbaz \inst{7}
     \and M. Still \inst{8}
     \and P. Wheatley \inst{3}}
     
\offprints{F. K. Sutaria, \email{F.K.Sutaria@open.ac.uk}}

\institute{The Open University, Milton Keynes, U.K. 
           \and University of Southampton, U.K.,
           \and X-ray Astronomy Group, Dept of Physics and Astronomy, 
            University of Leicester, University
Rd, Leicester, LE1 7RH, U.K.,
           \and SRON, Utrecht, The Netherlands, 
           \and IAC, La Laguna, Tenerife, Spain,
           \and Institute of Astronomy and Astrophysics, National Observatory of Athens, 
           P.O. Box 20048 - Athens 11810, Greece,
           \and Dept. of Astrophysics, Oxford University, U.K.,
           \and NASA/GSFC, Maryland, U.S.A.}
           
\authorrunning{F. K. Sutaria et al.}
\titlerunning{XMM-Newton observation of GU Mus in quiescence.}
\date{Received ; Accepted }       

\abstract{ The soft X-ray transient GU Mus has been detected by XMM-Newton in the quiescent
state.  The source is very faint, with a 0.5-10.0 keV unabsorbed flux of $\simeq 1.1 \times
10^{-14}$ ergs cm$^{-2}$ s$^{-1}$.  The spectra is well fit by an absorbed powerlaw with a
photon index of $ \alpha = 1.6 \pm 0.4$, close to the value seen when the source was in the
low/hard state in Aug.  1991.  From our observed luminosity, it seems unlikely that the
quiescent state emission is dominated by coronal X-rays from the secondary.  The flux also
appears to be in agreement with the ADAF model of BH-transients in quiescence.
 }

\maketitle

% begin{center}

\section{Introduction:}

 %Physical reasons for looking at SXTs in quiescence.

Soft X--ray transients (SXTs) are compact binaries where a neutron star (NS) or
black hole (BH) primary accretes from a Roche--lobe filling donor star.  While there
exists unique observational signatures for the identification of a neutron star in
these systems, e.g.  through X--ray bursts which are thought to be due to
thermonuclear burning of superheated accreted matter on the surface of a neutron
star, there is no similar identifier for an event horizon -- the unique
characteristic of a BH-system.  A knowledge of the mass function of the
system (from optical observations of the companion), can at most give a lower limit
to the mass of the compact object.  In the case of stellar mass BHs, this is usually not
very definitive, given the present uncertainty on the limiting mass of neutron
stars.

The spectra of SXTs in outburst appear to be dominated by a radiatively efficient,
optically thick, geometrically thin, accretion disc in a hot, high viscosity
state.  However, spectra of BH-SXTs in quiescence do not appear to be a simple
accretion disc phenomenon alone (e.g..\ Lasota 2001).  A simple, two component
model for SXTs in quiescence consists of an outer, thin, accretion disc, which
serves as a reservoir for the accumulated mass.  An inner, hot,
advection-dominated accretion flow (ADAF) transports energy from the inner edge of
the accretion disc to the compact object (Menou et al.  1999).  In the case of
NS-SXTs, this would heat up the surface of the NS (without thermonuclear runaway)
causing it to radiate, or, in the case of BH systems, carry the energy, along with
matter, out of sight beyond the event horizon.  Thus the quiescent state spectra
of SXTs would yield information on the nature of the compact object.  Hence, in
recent years the quiescent properties of SXTs and their white dwarf analogues
(dwarf novae) have taken centre stage in the quest to prove the existence of black
holes by detecting the event horizon, and in the search for the holy grail of
accretion disc theory, the viscosity mechanism (Garcia et al.,2001, and Narayan et
al., 2001).

The two-component model discussed above is not without inconsistencies.
 Inclusion of viscous heating of electrons, along with that of protons,
 accounting for convectively unstable flows, and magneto--rotational instability
 in 3-D numerical computations of ADAFs now show that ADAFs develop strong
 outflows (Hawley et al.\ 2001).  Also, the large mass flow inherent in ADAF
 models predicts a much larger L$_{\rm quiesc}$ in NS systems than observed.
 However, all ADAF models consistently predict that BH-SXTs should have a much
 lower quiescent flux (by a factor of $ \simeq 100$) relative to the NS SXTs.

 Prompted by the apparent failure of ADAF for NS-SXTs, alternative explanations for the
quiescent emission have been put forward, including coronal emission from
companion stars (e.g..\ Bildsten \& Rutledge 2000; but these exceed the existing
quiescent BH SXT detections), and emission from the neutron star surface that
was heated in earlier outbursts (Brown et al.\ 1998).

So far, quiescent state detections, or upper limits, exist for 6 NS-SXTs and 8
BH-SXTs.  The quiescent luminosities of these systems in the 0.5-10.0 keV range
(Table 2, Narayan et al., 2001), normalised to $10^{38}$ ergs s$^{-1}$ are
plotted as a function of the orbital period $P_{\rm orb}$ in fig.  \ref{fig:  LvsP}.
Systems with similar orbital periods are expected to be at the same point in
their evolutionary history, and hence expected to have similar accretion rates
$\dot M$ (however, see Sect. \ref{sec: Discussion}).  
In accordance with the ADAF models, the BH-SXTs do appear to be
fainter than the NS SXTs.  We discuss the cause of this scatter in Sect.
\ref{sec:  Discussion}.  We had designed our observation of the BH-SXT GU Mus in
order to (a) investigate the ADAF prediction of the relative faintness of
quiescent BH systems over NS systems with similar orbital period and (b) to
obtain a potential spectral signature for the existence of an event horizon --
i.e., the search for the indisputable evidence of a black hole.  GU Mus is an
ideal candidate for testing these hypotheses because no X-ray activity has been
detected from it since the last outburst in Jan.  1991.  We discuss the outburst
properties, the present observation and our results in the following sections.

\begin{figure}[h]
 \resizebox{\hsize}{!}{\includegraphics{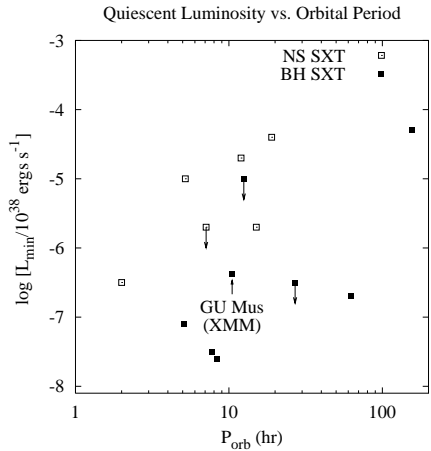}}
 \caption{Quiescent luminosities of BH-- and NS-- SXTs plotted against their
 orbital periods. The data is taken from Table 1 of Narayan et al. 2001. All
 luminosities are in the 0.5 to 10 keV band.}
 \label{fig: LvsP}
 \end{figure}
 
\section{Earlier Observations of SXT GU Mus (Nova Muscae 1991):}

  In X-rays, GU Mus (also known as GS 1124-68, Nova Mus, Nova Muscae 1991) was
initially detected as a bright SXT in outburst by GRANAT-WATCH and the Ginga-ASM
detectors, between Jan 9 to 11, 1991 ( Lund et al.  1991 and Makino, 1991), and by
the ROSAT-All Sky Survey during Jan 24/25 1991.  The combined high/soft state
ROSAT and Ginga spectrum was fitted by a two component model, consisting of a
multitemperature disk blackbody (Mitsuda et al.  1984), with a maximal disk
temperature of 0.96 keV plus a hard powerlaw component with a photon index of $
\simeq 2.45 $ (Greiner et al.  1994).  Subsequent photometric observations of the
optical counterpart during April '92 to July '94 revealed a periodicity of $\sim
10.38$ hr.  in this source (Orosz et al.  1996).  From optical spectroscopic
measurements in quiescence (Orosz et al 1996, and references therein), the mass
function for the system was measured at $f(M)=3.01 \pm 0.15 M_{\odot}$, and limits
on the inclination angle were set at $ 54\degr \le i \le 65\degr$.  The mass of
the compact object and its companion are $3.86 \pm 0.19 (\sin^{-3}i)$ $M_{\odot}$,
and $0.51 \pm 0.06 (\sin^{-3}i)$ $M_{\odot}$ respectively.  Infrared observations
(Gelino et al.  2001) set narrower limits on the inclination angle $i = 54\degr
\pm 1.5\degr $, and restrict the mass of the BH to $6.95 \pm 0.6 M_{\odot}$.

 Distance estimates for GU Mus range from 2.8 kpc to $ \ge 5.5 $ kpc,
depending on the mass of the compact object, the spectral type of the companion,
and the accretion model.  Optical photometric observations of the companion,
revealed it to be a main sequence dwarf star with spectral type K3-5 V, thus
implying a distance of $5.5 \pm 1$ kpc.  Infrared observations in 1995 (Shahbaz
et al., 1995) of GU Mus give a distance limits of 2.8 kpc to 4.0 kpc.  Finally,
combined X-ray Ginga and ROSAT observations of GU Mus in outburst yielded an
X-ray column density $N_H = 2.6 \times 10^{21} \rm cm^{-2}$, assuming a thin
disc model for the spectra. This was consistent with
the extinction obtained from both IUE ($E_{B-V} \simeq 0.2-0.3$), and HST
observations ($E_{B-V}= 0.287 \pm 0.004 $, Cheng et al., 1992), and is larger
than the expected galactic reddening in this direction, suggesting that the GU
Mus transient is located behind the galactic disc (Greiner et al. 1994). 
Because the estimate of distance from X-ray observations is highly model 
dependent, we use here the "optical value" of $d=5.5$ kpc as our reference distance.

Prior to 1991, there were no outbursts detected from this source (Greiner et al.  '94).  Following
the 1991 outburst, the source luminosity showed an exponential decline typical of BH-systems, with
a characteristic timescale of 21.9 days.  Though there was a second luminosity increase $ \sim 80$
days after the outburst, and again at $\sim 200$ days, the source showed an overall steady decline
in luminosity, and it fell below the the Ginga/LAC detection limits (0.3 mCrab) 282 days after the
outburst (Ebisawa et al.  1994).  It was not detected in the ROSAT (0.3-2.4 keV) band $ \sim 410$
days after outburst, and the source flux had fallen below $\simeq 5.0 \times 10^{-14}$ ergs
cm$^{-2}$ s.  It has remained in the quiescent state ever since, with no significant detection
reported until now.

\section{The XMM-Newton observation:}

GU Mus was observed by XMM-Newton detectors on Feb 24-25, 2001, with an 
exposure time of $\sim 34100$s.  We used pipeline processed data for our analysis,
because a comparison of the calibration files used in the pipeline processing
with those used at the time of analysis suggested that no significant
improvements would have resulted with any recalibration of the data.  The events
in the EPIC-MOS1, EPIC-MOS2 and EPIC-PN datasets were filtered using the
SAS-xmmselect task, using the event selection criteria recommended in the
``Users Guide to the XMM-Newton Science Analysis System" (2001).  The total
counts in the three EPIC CCDs from GU Mus was $\sim 100$.  Both EPIC-MOS and
EPIC-PN observations were carried out in the thin filter mode.

Among the various sources detected on the EPIC CCDs, we identified GU Mus from
its optical coordinates (RA $= 11^{h} 26^{m} 26\fs65 $ and Dec.$= -68\degr 40
\arcmin 32\farcs2 $, J2000) (Della Valle et al., 1991).  From a time series
analysis of the entire EPIC-PN, MOS1 and MOS2 data, we find that there are times
of large, random fluctuation over the entire field of view of both EPIC-PN and
MOS CCDs.  This high background rate over almost 50\% of the exposure time was
possibly due to detection of soft protons, or other high energy particles by the
EPIC-CCDs.  By restricting our analysis to the intervals of low background, our
effective exposure time was reduced to $\sim 15800$ s.

The data was binned and analysed using the XSPEC (version 11.1) package.  The response
and auxilary (rmf- and arf- ) files were generated for each CCD using the SAS tasks "rmfgen"
and "arfgen" respectively.  We extracted the background spectrum from a region of the chip
which was free of any significant sources, and which was on same section of the CCD as the
source.  For EPIC-PN, in the 0.3 to 14 keV range, the countrate for the background spectra
was $9.4(\pm 2.4) \times 10^{-4} \rm s^{-1}$ and the background subtracted countrate in the
source region was $1.64(\pm 0.45) \times 10^{-3} \rm s^{-1}$.  Though faint, the source
stands well above the statistical fluctuation of the background.  The background subtracted
countrates in EPIC-MOS1 (0.3-12 keV) was $0.72(\pm 0.44) \times 10^{-3}$ s$^{-1}$ and in
EPIC-MOS2 (0.3-12 kev) it was $0.77(\pm 0.57)\times 10^{-3}$ s$^{-1}$. The count rate in
the MOS1 and MOS2 detectors is too low to give useful spectra even after binning, so since
the MOS1 and MOS2 have similar response functions, their spectra were grouped together and
analysed.  The EPIC-PN spectra are analysed separately.

For both EPIC-PN and EPIC-MOS, because of the very low number of counts, a meaningful fit was
only possible if the column density was kept fixed.  We chose $N_H = 2.6 \times 10^{21} \rm
cm^{-2}$ -- as obtained from fits to ROSAT and Ginga data (see above).  The EPIC-PN data were
binned so as to produce at least 8 counts/bin, and the MOS1 and MOS2 data were binned to at
least 5 counts/bin.  Since $\chi^2$ fitting requires about 30 counts/bin to be meaningful, we
used the Cash-statistic (XSPEC C-statistic, see Cash 1979) here instead, which puts better
limits on the fitting parameters (Table \ref{tab:  EPIC-PN}).  XSPEC version 11.1, used
here, allows C-statistic fits on data with the background spectra read in.  The C-statistic is
defined as the logarithm of the probability that the number of counts in a given phase bin is
actually reproduced by the fitted model.  Since C-stat does not provide a goodness of fit, for
the purpose of comparison, we computed both the goodness of fit from Monte-Carlo probability
calculations (Col. 5, Table \ref{tab:  EPIC-PN}), as well as the $\chi^2$ statistic.  The
"goodness of fit" should hover around 50\% if the observed data was produced by the fitted
model.  The chi-statistic was calculated for each fit using the parameters obtained from
C-statistic fitting and quoted in the Table.  Fig.\ref{fig:  GUMus_pn_freezeNH} shows the
EPIC-PN spectrum fitted using an absorbed power-law model and C-statistic. The $\chi^2$-deviation
displayed in the figure was calculated using the C-stat fitted parameters.

\begin{figure}[h]
\rotatebox{270}{\resizebox*{7.0cm}{!}{\includegraphics{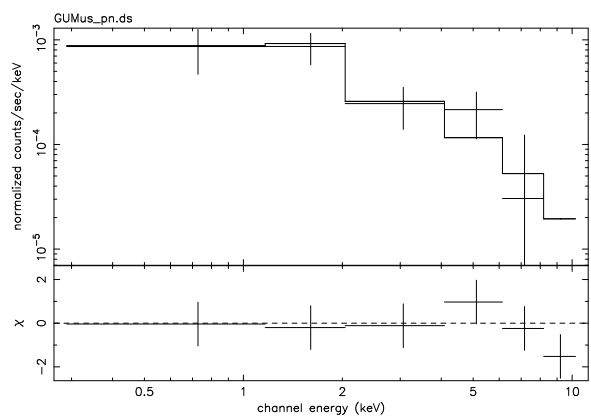}}}
  \caption{ EPIC-PN spectrum and the fitted model (see Sect. 3) of GU Mus in 
quiescence.
  The absorbed powerlaw model is fitted here, with  $\log N_H$ fixed at 21.4.
  The residuals are plotted in the lower panel.
  }
  \label{fig: GUMus_pn_freezeNH}
\end{figure}

In addition to the absorbed power-law model, we have also fitted our data to the absorbed
bremsstrahlung, Raymond-Smith (solar abundance) and black-body models.  The EPIC-PN spectra
were fitted over the 0.2-12 keV range, and the results for each model are presented in Table
\ref{tab:  EPIC-PN}.  In order to facilitate comparison with the Chandra observations of the
spectra of other BH-SXT (Kong et al., 2002), we quote the (absorbed) flux in the 0.3-7 keV
range as well as in the 0.3-12 keV range in Table \ref{tab:  EPIC-PN}.  The fits to the
EPIC-PN spectra suggests that while the $\chi_{\nu}^2/{\rm d.o.f.}$ are comparable for all models
used, the blackbody model can at least be counted out, because of the very low value of
absorption.  Similarly, the bremsstrahlung and the coronal emission model (Raymond-Smith)
also constrain the parameter values very poorly, so we are inclined to treat the absorbed
powerlaw as the model that best describes the (admittedly sparse) data. Using the 90\%
confidence limits on the powerlaw fit with constant $\alpha$ (see fig.  \ref{fig:
contours}), the unabsorbed, $0.5-10$ keV flux is $1.1^{+1.1}_{-0.4} \times 10^{-14}$ ergs
cm$^{-2}$ s$^{-1}$.  The energy range $0.5-10.0$ keV is being used to allow comparison with
the earlier (mainly Chandra) observations plotted in fig.  \ref{fig:  LvsP}.  Since the
distance to GU Mus is uncertain, to estimate the luminosity we use a reference distance $d =
5.5 \rm kpc$, as obtained from optical observations (Orosz et al.  1996).  Thus the
unabsorbed EPIC-PN flux quoted above implies a luminosity of $4.0 \times 10^{31} (d/5.5 \
{\rm kpc})^2$ ergs s$^{-1}$ in the 0.5-10 keV range.

\begin{figure}[h]
\rotatebox{270}{\resizebox*{7.0cm}{!}{\includegraphics{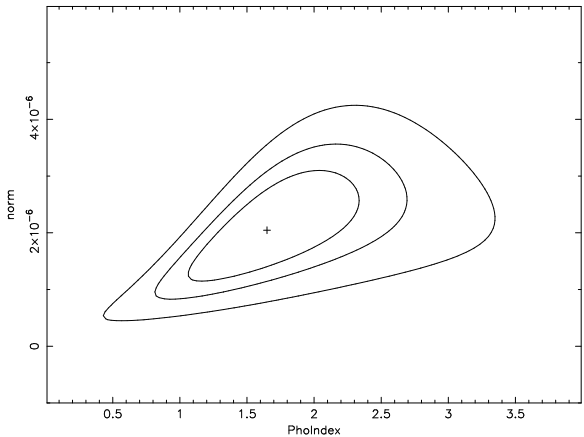}}}
  \caption{ The contours plotted above are for  68.3\%, 90\% and 99\% confidence limits
on the values of the photon index $\alpha$ and powerlaw normalisation $\it norm$, 
assuming that $N_H = 2.6 \times 10^{21}$ cm$^{-2}$.}
  \label{fig: contours}
\end{figure}

\begin{table*}[h] 
\caption{Best fit spectral parameters for EPIC-PN. $kT$ represents 
the plasma temperature or the black body temperature, depending on the model.
All uncertainties are quoted in 90\% confidence limit. The absrobed flux is quoted here.
 \label{tab:
EPIC-PN}} 
\begin{tabular}{cccccccc} 
\hline 
\hline
Model & $N_H$ & $\alpha$ & $kT$ & CASH & $\chi_{\nu}^2/{\rm d.o.f.}$ & $f|_{0.3-7.0}$ & $f|_{0.3-12.0}$ \\
\hline
 & $10^{21}$ & & & M-C & & $ \times 10^{-14}$ & $ \times 10^{-14}$\\
 & cm$^{-2}$ & & keV & Prob. &  & ergs cm$^{-2}$ s$^{-1}$ & ergs cm$^{-2}$ s$^{-1}$\\
\hline 

Power-law & $2.3^{+8.9}_{-2.3}$ & $1.6^{+1.6}_{-0.9}$ & -- & 44.8 \% & 1.12/3 & $0.90$ & $1.31$ \\
& $2.6$ (const.)  & $1.6^{+0.8}_{-0.6} $ & -- & 44.9 \% & $0.85/4$ & $0.93$ & $1.34$ \\

Bremsstrahlung & $1.9^{+0.1}_{-1.9}$ & -- & $ 9.7^{+190.2}_{-8.2}$ & 38.4 \% & 1.05/3 & $0.91$ & $1.12$ \\
               & $2.6$ (const.)& -- & $7.4^{+192.6}_{-5.6}$ & 39.9 \% & $0.82/4$ & $0.88$ & $1.12$\\

Raymond-Smith & $1.8^{+7.4}_{-1.8}$ & -- & $11.1^{+52.9}_{-9.5}$ & 45.8 \% & 1.11/3 & $0.96$ & $1.30$ \\
              & $2.6$ (const.) & -- & $6.3^{+57.7}_{-4.5}$ & 43.8\% & $0.87/4$ & $0.97$ & $1.20$ \\

Black body & ${1.6 \times 10^{-4}}^{+3.9}_{-1.6 \times 10^{-4}}$ & -- & $0.88^{+0.48}_{-0.40}$ & 49.5 \% 
                 & $1.35/3$ & $0.87$ & $0.91$ \\
          & $2.6$ (const.) & -- & $0.83^{+0.52}_{-0.46}$ & 67\% & $1.5/4$ & $0.84$ & $0.87$ \\

\hline
\end{tabular}
\end{table*}

The MOS1 and MOS2 data were fitted simultaneously, allowing only the relative normalisations
to vary.  Fitting the 0.3-10 keV spectra, and using an absorbed powerlaw model, we find that
C-statistic fit parameters are reasonable only if we keep $N_H$ constant.  The best fit
photon index was $\alpha = 1.43 \pm 0.52$, for $\chi_{\nu}^2/{\rm d.o.f.}  = 1.15/7$.  The absorbed
0.5-10.0 keV flux is $ 1.8 \times 10^{-14}$ and $ 1.2 \times 10^{-14}$ ergs cm$^{-2}$
s$^{-1}$ in EPIC- MOS1 and MOS2 respectively.  The other three models quoted in Table
\ref{tab:  EPIC-PN} do not fit the data well, resulting in unrealistic and badly constrained
values of the temperature parameter in each model, and hence are not reported here.

 The source was too weak to be detected in the RGS1 and RGS2 spectrometers, nor was it detected in
the UVM2 and UVW1 filters by the optical monitor.  For an exposure time of 2860 s, the limiting count
rate for detection in the UVW1 filter was $2.4 \times 10^{-2}$ s$^{-1}$, and in the UVM2 filter, it
was $1.36 \times 10^{-2}$ s$^{-1}$.

\section{Discussion} 
\label{sec: Discussion}

Our observation had been planned with the intention of using the spectral
characteristics of GU Mus to provide an observational discriminant between NS-
and BH-SXTs.  However, we find that, while we have a clear detection of GU Mus
in the quiescent state, at $\sim 100 $ counts, the spectra have insufficient
counts to constrain well even the simple absorbed power-law model.  By fixing
the column density at $N_H = 0.26 \times 10^{22}$ cm$^{-2}$, the fit to the
EPIC-PN spectra constrains the photon index to $\alpha = 1.6 \pm 0.4$.  This
appears to be consistent with both the outburst and post-outburst observations,
where the photon index of the hard component decreased from its high/soft state
value of 2.2-2.7 (Jan.  to Apr.  1991) to its low/hard value of $\sim 1.6$ in
Sept.  1991 (Ebisawa et al.  1994).  Besides GU Mus, only 5 other BH-SXTs have
been actually detected in quiescence (for Chandra observations, see Kong et al.
2001, and Garcia et al.  2001) and these luminosities have been plotted in fig.
\ref{fig:  LvsP}, as a function of orbital period $P_{\rm orb}$.  In addition, upper
limits for the BH SXTs 4U 1543-47 ($P_{\rm orb} = 27$ hr., Orosoz et al.  1998) and
H1705-250 ($P_{\rm orb} = 12.7 $ hr, Narayan et al, 1997) have also been plotted.
%Distance estimated used are based on quiescent stage optical classification of 
%the secondary .  
The luminosities for the Chandra observations included
in fig.  \ref{fig:  LvsP}, were calculated (Narayan et al., 2001 and references
therin) in the energy range 0.5-10.0 keV, by fitting to a power-law spectrum
with photon index $\alpha \sim 2.0$ and column densities consistent with the
optical extinction.

The evaluation of the intrinsic X-ray luminosity of GU Mus in quiescence is
uncertain, chiefly because of the poor constraints on the distance.  In fig.
\ref{fig:  LvsP}, we have plotted the value of $\log L_{quies}/10^{38} = -6.40 $
for GU Mus, using our reference value of 5.5 kpc.  
%This increases to -5.8 if we take the distance to be 11 kpc, evaluated from 
%X-ray observations in outburst (Greiner et al.  1994).  
In terms of the Eddington luminosity, using an
inclination of $i= 60^{\circ}$, $L_{0.5-10.0}/L_{Edd} = 9.25 \times 10^{-8}
(d/5.5 \rm kpc)^2$.  We plot the luminosities in fig.  \ref{fig:  LvsP} as
normalised to a factor of $10^{38}$ ergs s$^{-1}$, instead of the ratio
$L_{\rm quiesc}/L_{ \rm Edd }$, as preferred by others (e.g.  Garcia et al 2001), because
the Eddington luminosity $L_{ \rm Edd }$ will be $ \sim 10$ times larger in BH
systems than in NS systems, thus leading to an artificial separation between the
BH and NS systems.  A natural separation can be expected in compact binaries
with P$\leq 12$~hrs when the accretion rate $\dot M $ is driven by orbital
angular momentum losses $\dot J$.  If these are dominated by magnetic braking,
$\dot J$ is the same for identical donor stars, so that $\dot M $ is
$\propto 1/J$ i.e.\ roughly to $1/M_{\rm primary}$ (e.g.\ King et al.\ 1997).
Hence $\dot M$ in BH systems would be expected to be smaller than in NS systems.
Normalisation relative to $L_{\rm Edd}$ is reasonable only if the event were driven
by gravitational radiation, since, in that case, $\dot M$ would scale with $M$.
However, the truth may be more complicated.  Some short--period NS systems may
be remnants of a thermal--timescale mass transfer phase (King et al.\ 2001), and
at least one short--period BH system (XTE J1118+480) shows evidence
nuclear--processed material (Haswell et al.\ 2001), a clear sign of an evolved
donor. The secondary in the BH-SXT GS 2000+25, too, has been found to be 
an evolved donor (Harlaftis et al, 1996).  All this suggests a greater spread of
$\dot M$ at a given $P_{\rm orb}$.

An alternative explanation for SXT quiescent X-ray emission was given by
Bildsten and Rutlidge (2000), who suggested that it arose from the corona of the
(necessarily) rapidly rotating companion star.  In order to investigate this
possibility, we fitted our data to the Raymond-Smith model for a coronal plasma,
assuming solar abundances.  If the emission is indeed coronal in origin, rather
than an ADAF in the inner disk region, then the temperature of the plasma should
be $ \le 1.4 $ keV (Dempsey et al.  1993), and the ratio of the soft X-ray to
bolometric luminosities be $ L_{\rm SX}/L_{ \rm bol} < 10^{-3} $.  While our fits
constrain the temperature too poorly to compare, we do note that the {\it
absorbed} flux in the 0.4-2.4 keV range is $3.3 \times 10^{-15}$ ergs cm$^{-2}$
s$^{-1}$, (EPIC-PN spectra with constant $N_H$) which implies, $L_{\rm  SX}/L_{ \rm bol} >
8.96 \times 10^{-3}$.  In order to calculate $L_{\rm bol}$, we used $V = 20.5$ for
GU Mus, with a $50 \%$ contribution from the accretion disk (Orsoz et al.
1996), resulting in $V_{\rm secondary} = 21.25$.  The other parameters were
$E(B-V)=0.3$ ( Della Valle et al.  1991), $A_V = 3.1 E(B-V)$ and distance $d=
5.5$ kpc.  Thus, based on luminosity estimates alone, it maybe unlikely that the
quiescent state X-ray emission originates from the companion star's corona.  We
also note, however, that of all the other BH-SXTs observed in quiescence so far
(GRO J0422+32, A0620-00, XTE J1550-564, V404 Cyg, GRO J1655-40 and GS 2000+25,
Kong et al.  2000), the stellar corona appears to contribute a major portion of
the soft X-ray luminosity only in GS 2000+25, although as yet no spectra exists
for the quiescent state of this source.

%\begin{figure}
%\rotatebox{270}{\resizebox{\hsize}{!}
%{\includegraphics{Mos1Mos2_frzNH_const.ps}}}
%   \caption{ EPIC-MOS1 and MOS2 spectra of GU Mus in quiescence. 
%The spectra were grouped together and fitted, with $\log N_H$ being kept 
%fixed at 21.41, and allowing only relative normalisations to vary.}
%  \label{fig: GuMus_M1M2_frzNH}
%\end{figure}
\end{document}